\documentclass[aps,prd,nofootinbib,superscriptaddress,onecolumn,notitlepage,preprintnumbers,fleqn]{revtex4-1}

\usepackage{amsmath,graphicx,verbatim,epsfig,amssymb,dsfont}
\usepackage{epstopdf}
\usepackage[latin1]{inputenc}
\usepackage{amsmath,graphicx}
\usepackage{simplewick}
\usepackage{datetime}
\usepackage[colorlinks=true,linkcolor=blue,citecolor=blue]{hyperref}
\usepackage[usenames,dvipsnames]{color}
\usepackage{setspace}
\usepackage[mathscr]{eucal}

\newcommand{\eps}{\varepsilon}

\newcommand{\Ga}{\Gamma}

\newcommand{\bn}{{\bar n}}

\newcommand{\veuv}{\varepsilon_{\rm {UV}}}

\newcommand{\be}{\begin{equation}}
\newcommand{\ee}{\end{equation}}
\newcommand{\bea}{\begin{eqnarray}}
\newcommand{\eea}{\end{eqnarray}}
\newcommand{\balign}{\begin{align}}
\newcommand{\ealign}{\end{align}}
\newcommand{\as}{\alpha_s}

\newcommand{\cd}{\!\cdot\!}

\newcommand{\mean}[1]{\left< #1\right >}

\newcommand{\bg}{\begin{gather}}
\newcommand{\foma}{\end{gather}}

\newcommand{\noopsort}[1]{}

\newcommand{\vecb}[1]{\mbox{\boldmath $#1$}}
\newcommand{\vecbe}[1]{\mbox{\boldmath ${\scriptstyle #1}$}}

\def\<{\langle}
\def\>{\rangle}

\def\g{\gamma}  \def\G{\Ga}
\def\d{\delta}  \def\D{\Delta}
\def\l{\lambda}   
\def\s{\sigma}
  
\def\x{\xi}

\def\m{\mu}

\def\({\left(}
\def\[{\left[}
\def\){\right)}
\def\]{\right]}

\def\ln{\hbox{ln}}

\def\bk{\bar k}

\newcommand{\ben}{\begin{eqnarray}}
\newcommand{\een}{\end{eqnarray}}

\newcommand{\bef}{\begin{figure}[htb]\centering}
\newcommand{\eef}{\end{figure}}

\usepackage[normalem]{ulem} 
\renewcommand\sout{\bgroup \color[rgb]{1,0,0} \ULdepth=-.5ex \ULset}

\begin{document}

\title{
Proper definition and evolution of generalized transverse momentum dependent distributions}

\author{Miguel G. Echevarria}
\email{mgechevarria@icc.ub.edu}
\affiliation{Departament de F\'isica Qu\`antica i Astrof\'isica and
Institut de Ci\`encies del Cosmos, Universitat de Barcelona,\\
Mart\'i i Franqu\`es 1, 08028 Barcelona, Spain}

\author{Ahmad Idilbi}
\email{ahmad.idilbi@wayne.edu}
\affiliation{Department of Physics,
Wayne State University, 
Detroit, MI, 48202, USA}

\author{Koichi Kanazawa}
\email{koichi.kanazawa@temple.edu}
\affiliation{Department of Physics, SERC, 
Temple University, Philadelphia, Pennsylvania
19122, USA}

\author{C\'edric Lorc\'e}
\email{cedric.lorce@polytechnique.edu}
\affiliation{Centre de Physique Th\'eorique, \'Ecole polytechnique, 
CNRS, Universit\'e Paris-Saclay, F-91128 Palaiseau, France}

\author{Andreas Metz}
\email{metza@temple.edu}
\affiliation{Department of Physics, SERC, 
Temple University, Philadelphia, Pennsylvania
19122, USA}

\author{Barbara Pasquini}
\email{barbara.pasquini@pv.infn.it}
\affiliation{Dipartimento di Fisica, Università degli Studi di Pavia, I-27100 Pavia, Italy}
\affiliation{Istituto Nazionale di Fisica Nucleare, Sezione di Pavia, 
I-27100 Pavia, Italy}

\author{Marc Schlegel}
\email{marc.schlegel@uni-tuebingen.de}
\affiliation{Institute for Theoretical Physics, Tübingen University, 
Auf der Morgenstelle 14, 72076 Tübingen, Germany}




\begin{abstract}
We consider one of the most fundamental sets of hadronic matrix elements, namely the generalized transverse momentum dependent distributions (GTMDs), and argue that their existing definitions lack proper evolution properties. 
By exploiting the similarity of GTMDs with the much better understood transverse momentum distributions, we argue that the existing definitions of GTMDs have to include an additional dependence on soft gluon radiation in order to render them properly defined. 
With this, we manage to obtain the evolution kernel of all (un)polarized quark and gluon GTMDs, which turns out to be spin independent.
As a byproduct, all large logarithms can be resummed up to next-to-next-to-leading-logarithmic accuracy with the currently known perturbative ingredients.




\end{abstract}

\maketitle


\section{Motivation}
\label{sec:motivation}

It has been known for quite some time that hadronic matrix elements such as generalized parton distributions \cite{Mueller:1998fv,Ji:1996ek,Radyushkin:1997ki} (GPDs) and transverse momentum dependent parton distribution/fragmentation functions \cite{Collins:2011zzd,GarciaEchevarria:2011rb,Echevarria:2012js} (TMDs in general) are indispensable objects for studying fundamental properties of hadrons.  
With them one can probe nucleon tomography by investigating their spin and three-dimensional momentum distributions.

More recently a new class of hadronic matrix elements was introduced, which generalizes both GPDs and TMDs. 
In \cite{Meissner:2008ay}, correlation functions which describe off-forward scattering amplitudes were introduced, where the bi-local partonic fields are separated in all three light-front coordinates: ($z^+, z^- ~\rm{and}~ \boldsymbol{z}_\perp$). 
Given some kinematics imposed by an \emph{underlying} scattering process, one can consider a restricted class of matrix elements for which either $z^+=0$ or $z^-=0$. 
In this case the matrix elements reduce to what is known as \emph{generalized transverse momentum dependent parton distribution functions} (GTMDPDFs or simply GTMDs) \cite{Meissner:2008ay,Meissner:2009ww,Lorce:2013pza} (see also \cite{Ji:2003ak,Belitsky:2003nz}, and a recent review in \cite{Diehl:2015uka}). 
These hadronic quantities are off-forward matrix elements with explicit dependence on the longitudinal and the transverse momentum components of partons inside hadrons. 
As such, they are hybrid constructs of both classes of functions: GPDs and TMD bi-local correlators.
Notice that here we make a distinction between TMDs and TMD correlators, where the latter, also called \emph{unsubtracted TMDs}, are ill-defined due to uncanceled spurious rapidity divergences (see discussions in, e.g., \cite{Collins:2011zzd,GarciaEchevarria:2011rb,Echevarria:2012js}).

In this work we argue that the hadronic matrix elements called in the literature \emph{GTMDs}, as currently formulated and analyzed, are improperly defined. 
The last assertion results from the observation that these matrix elements lack proper evolution properties with respect to the renormalization scale $\mu$ and the rapidity scale $Q$. 
Also their operator product expansion into generalized parton distributions breaks down even for large enough transverse momentum, contrary to what the case should be.
Definitely these facts severely limit the predictive power of the underlying theory: QCD.
The upshot is that such matrix elements, although formulated from the basic QCD fields and being gauge invariant, cannot be admissible in the group of \emph{genuine QCD hadronic quantities}, such as the integrated PDFs, GPDs or the more recently reformulated TMDs.

When calculated perturbatively, as we show below for an unpolarized quark target, the existing definition of GTMDs suffers from unwanted rapidity divergences (RDs).
Those divergences prohibit the derivation of any evolution kernel: neither the anomalous dimension governing the running with respect to the renormalization scale $\mu$ nor the resummation of large logarithms~\footnote{This resummation allows us to obtain the $D$ term \cite{Echevarria:2012pw} (also referred to as $\tilde{K}$ in Collins' formulation \cite{Collins:2011zzd}) for TMDs.}. 
These spurious divergences appear in all perturbative calculations in the soft and collinear limits of QCD. Those limits are exactly where all partonic matrix elements live at,  and this occurs on a Feynman diagram-by-diagram basis. 
In those limits one obtains two  different kinds of Wilson lines which in turn cause the appearance of the spurious divergences.
As such, RDs appear in various Feynman diagrams contributing to the integrated PDFs, GPDs, TMD correlators and GTMD correlators. 

As it is well known, divergences have to be regularized. For RDs, and regardless of the methods implemented to regularize them, they do cancel for transverse momentum integrated matrix elements, such as PDFs and GPDs, when all the relevant contributions are added together \cite{Collins:1981uw} and this cancellation occurs order by order in perturbation theory. 
However for matrix elements with a specified transverse momentum dependence (or unintegrated quantities), whether in coordinate or momentum space, the cancellation is not complete among contributions from different classes of Feynman diagrams and the remaining rapidity divergences spoil any attempt for a three-dimensional hadronic description, at least pertubatively.

The existence of RDs for the currently formulated GTMDs should not be that surprising. 
Actually the situation is similar to the case of the by-now familiar TMDs \cite{GarciaEchevarria:2011rb} (see \cite{Echevarria:2015uaa} for gluon TMDs).
The fact that the GTMD correlator is off-diagonal in hadron momenta only makes the QCD corrections more laborious to obtain, however the fundamental observation regarding the non-cancellation of RDs remains the same as in the forward limit case.

At the operator level~\footnote{A generic vector $v^\m$ is decomposed as $v^\m=v^+\frac{n^\m}{2}+v^-\frac{\bn^\m}{2}+v_\perp^\m=(v^+,v^-,\vecb v_\perp)$, where $v^+\equiv \bn\cd v$ and $v^-\equiv n\cd v$, with the light-cone vectors defined by $n=(1,0,0,1)$ and $\bn=(1,0,0,-1)$. 
We also use $v_T\equiv|\vecb v_\perp|$, so that $v_T^2=\vecb v_\perp^2=-v_\perp^2>0$.}, the already existing definition of leading-twist quark GTMDs is given by \cite{Meissner:2009ww}
\begin{equation}\label{eq:GTMDunsub}
\phi_{\lambda\lambda^{\prime}}^{[\G],q} =
\frac{1}{2}\langle p^{\prime},\lambda^{\prime}|\,
\bar{q}(-z/2)\,\mathcal{W}_{n}(-z/2)\,
\G\,
\mathcal{W}_{n}^{\dagger}(z/2)\, q(z/2)\,|p,\lambda\rangle\Big|_{z^{+}=0}
\,,
\end{equation}
where $\lambda$, $\lambda^{\prime}$ are the nucleon helicities and the matrix $\G=\{\g^+,\g^+\g_5,i\s^{j+}\g_5\}$ stands for an unpolarized, longitudinally polarized or transversely polarized quark, respectively. 
Gauge invariance among regular gauges is satisfied by the inclusion of the collinear gauge link or Wilson line $\mathcal{W}_{n}$.
This Wilson line can be past-pointing or future-pointing, and the GTMDs depend in principle on this choice \cite{Lorce:2013pza}.
In this work, for definiteness, we choose the one consistent with DIS kinematics, however our observations and conclusions are valid in both cases. 
For DIS kinematics we have:
\begin{equation}\label{eq:WilsonLineDef}
\mathcal{W}_{n;\alpha\beta}(z)=
\left\{
\mathcal{P}\exp\Bigg[
-ig\int_{0}^{\infty}ds\ \bar{n}\cd A(z+s\bar{n})
\Bigg]
\right\}_{\alpha\beta}
\,,
\end{equation}
where the gluon field $A$ stands for collinear gluon field in the $n$ direction. To ensure gauge invariance among singular and regular gauges, one needs to also introduce  transverse gauge links at light-cone infinities ($z^-=\infty$),
see more details in~\cite{Belitsky:2002sm,Idilbi:2010im,GarciaEchevarria:2011md}.
Formally speaking, it is clear that when the states $\langle p^{\prime}, \lambda^{\prime} \vert$ and $\vert p,\lambda \rangle$ are taken with the same momenta we recover the standard TMD correlator, and when the transverse momentum dependence is integrated over we recover the standard GPDs.
It should be emphasized that the last statement holds, perturbatively, only for bare quantities. 
For renormalized GTMDs and GPDs one has to rely on the operator product expansion --whenever that expansion is valid-- to relate the \emph{physical} GTMDs and GPDs.

We show below that the basic matrix element in \eqref{eq:GTMDunsub} (and thus all the quantities derived from it) lacks any meaningful evolution properties. 
This is true for any pair of $\langle p^{\prime}, \lambda^{\prime} \vert$ and $\vert p,\lambda \rangle$ and also for gluon matrix elements (see \cite{Lorce:2013pza} for a generalization of \eqref{eq:GTMDunsub} to gluon case). 
In the next section we illustrate this explicitly for the first time at next-to-leading order in $\alpha_s$.
Thus, the way to proceed is to introduce a definition which does not suffer from any complications resulting from the three-dimensional dependence on the one hand, and allows for the recovery of the TMDs when the forward limit is taken on the other.
Motivated by the treatment of TMDs, we arrive at the following result for the \emph{properly defined quark GTMDs}:
\begin{equation}
\label{eq:GTMDPDFDef}
W_{\lambda\lambda^{\prime}}^{[\G],q}=
\frac{1}{2} \int\frac{dz^-d^2z_\perp}{(2\pi)^3}\,
e^{+i\(\frac{1}{2}z^-\bk^+-\vecbe z_\perp\cdot\vecbe\bk_\perp\)}\,
\phi_{\lambda\lambda^{\prime}}^{[\G],q}(0,z^-,\vecb z_\perp)\,
S^{\frac{1}{2}}(z_T)\, ,
\end{equation}
where the soft function~\footnote {In principle the soft function has to include transverse gauge links $T_{sn(s\bn)}$ \cite{GarciaEchevarria:2011md}.}
\begin{align}\label{eq:sf}
S(z_T)&=
\frac{\mathrm{Tr}_c}{N_{c}}
\langle0|
\mathcal{S}_{n}^{\dagger}\(-\frac{z}{2}\)\,\mathcal{S}_{\bar{n}}\(-\frac{z}{2}\)\,
\mathcal{S}_{\bar{n}}^{\dagger}\(\frac{z}{2}\)\,\mathcal{S}_{n}\(\frac{z}{2}\)|0\rangle\Big|_{z^{\pm}=0}
\,,
\end{align}
and for DIS kinematics and for arbitrary point $z$, the soft Wilson  lines are given by
\begin{align}\label{eq:WilsonLineDef2}
\mathcal{S}_{n;\alpha\beta}(z) &=
\left\{
\mathcal{P}\exp\Bigg[
ig\int_{-\infty}^{0}ds\ n\cd A(z+sn)
\Bigg]
\right\}_{\alpha\beta}
\,,\qquad
\mathcal{S}_{\bn;\alpha\beta}(z) =
\left\{
\mathcal{P}\exp\Bigg[
-ig\int_{0}^{\infty}ds\ \bn\cd A(z+s\bn)
\Bigg]
\right\}_{\alpha\beta}
\,.
\end{align}
Note that in \eqref{eq:sf} we use the \emph{natural} directions of the Wilson lines which include both past- and future-pointing soft Wilson lines, which are dictated by DIS kinematics. 
They follow from the eikonalization of the Feynman propagators. 
In \cite{Collins:2004nx,Collins:2011zzd} it was argued that all Wilson lines of the soft factor could, for instance, be chosen future-pointing, which right away would make the universality of the soft function evident between DIS and electron-positron annihilation processes.
However there is no need for such an {\it a priory} choice. 
One can first derive factorization using the directions of the Wilson lines imposed by the underlying kinematics and then, in a second step, investigate the universality properties of the obtained contributions to the factorized observable, such as the soft function.
Universality studies of the soft function using explicit calculations were carried out recently in \cite{GarciaEchevarria:2011rb,Echevarria:2012js,Echevarria:2014rua,Echevarria:2015byo}. 
More specifically, a 1-loop analysis can be found in \cite{GarciaEchevarria:2011rb}, and the extension to two loops is given in \cite{Echevarria:2015byo}. 
It was also argued that these fixed-order results should generalize to all orders in perturbation theory \cite{Echevarria:2012js,Echevarria:2014rua,Echevarria:2015byo}. 
In all of these works, it was demonstrated explicitly that no matter which directions of the soft Wilson lines are chosen, universality is still maintained.

Few additional remarks are in order. 
First, the given new definition of quark GTMDs can be straightforwardly extended to the case of gluon GTMDs.
The collinear contribution $\phi$ is implicitly understood to be the \emph{pure} collinear contribution, in the sense that soft contamination, encountered perturbatively, is already subtracted out. 
For more discussion on this issue, see e.g. \cite{Echevarria:2014rua,Manohar:2006nz}.
The appearance of the \emph{square root} of the soft function is motivated by the fact that the soft function contains RDs from both collinear and anti-collinear regions in rapidity space (see Fig.~3 in \cite{Echevarria:2012js}).
Thus, in order to eliminate the contribution from the anti-collinear region to the collinear correlator $\phi$, we need to take just \emph{half} of the soft function. 
Here we are simplifying the presentation by assuming a symmetry between the two collinear sectors, which corresponds to $\alpha=1$ in \cite{Echevarria:2012pw}.

For the TMDs (not GTMDs), the factor $\sqrt{S}$ results from the splitting of the soft function contribution $S$ among two TMD correlators --representing two collinear sectors--, where $S$ and the correlators appear \emph{naturally} from a factorization statement for a given actual physical process. 
For GTMDs the situation is currently different, since there is not, so far, any (factorized) process to rely on.
Actually, it is a very important question if/how GTMDs can be related to experimental observables. 
In fact just recently it has been argued that diffractive dijet production in electron-proton collisions could give access to gluon GTMDs \cite{Hatta:2016dxp} in the small-$x$ region.
The importance of properly defining quark and gluon GTMDs, extracting their evolution kernels, motivating their lattice calculation and experimental measurements, relies not only of them being fundamental objects of QCD, but also on their  connection to orbital angular momentum of partons inside hadrons \cite{Lorce:2011kd,Lorce:2011ni,Hatta:2011ku,Hagler:2003jw,Kanazawa:2014nha,Rajan:2016tlg}.

\section{GTMDs at NLO: emergence and cancellation of rapidity divergences}
\label{sec:nlo}

\begin{figure}[t]
\begin{center}
\includegraphics[width=0.35\textwidth]{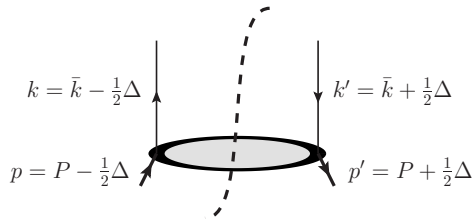}
\end{center}
\caption{\it
Kinematics for GTMDs in the symmetric frame.
}
\label{fig:gtmd}
\end{figure}

In this Section we present the perturbative calculation of the GTMDs for an unpolarized quark target, given in \eqref{eq:GTMDPDFDef} with $\G=\g^+$.
We use the $\delta$-regulator (see for instance  \cite{GarciaEchevarria:2011rb}) to regularize rapidity and infrared (IR) divergences~\footnote{In principle one should distinguish between the regulators that belong to $n$- and $\bn$-sectors, but for simplicity and without loss of generality we take them equal, i.e. $\tilde\D^\pm=\tilde\D$ and $\d^\pm=\d$. One should be careful not to confuse the regulator $\tilde\D$ with the four-momentum transfer $\D^\m$.}, and dimensional regularization in $\overline{\rm MS}$-scheme ($\m^2\to\m^2 e^{\g_E}/(4\pi)$) for ultraviolet (UV) ones. 
All the results below were obtained using Feynman gauge.

It is convenient to work in a symmetric frame, Fig.~\ref{fig:gtmd}, with the average nucleon momentum $P=\frac{1}{2}(p'+p)$, the \emph{momentum transfer} $\D=p^{\prime}-p=k'-k$ and the average quark momentum $\bk=\frac{1}{2}(k'+k)$.
These momenta are parametrized as 
$P^{\mu}=(P^{+},P^-,\vecb 0_{\perp})$ 
and 
$\D^{\m}=(-2\xi P^{+},2\xi P^{-},\vecb\D_\perp)$, with 
$P^-=\tfrac{\D_T^{2}+4M^{2}}{4(1-\xi^{2})P^{+}}$
and $M$ being the nucleon mass.
We work in a frame in which $p$ and $p'$ have very large \emph{plus} components.
The GTMDs $W_{\lambda\lambda^{\prime}}^{[\G],q}$ depend on the kinematical variables 
$(x,\xi,\bk_{T}^{2},\D_{T}^{2},\vecb \bk_\perp\cd\vecb\D_\perp)$,
with $x=\bk^+/P^+$ and $\x=-\D^+/(2P^+)$. 
They are also functions of the renormalization and rapidity scales, as we discuss below, and our NLO results are valid in the DGLAP region, i.e. $|\xi| < x$. 

Note that the generic structure of GTMDs is also constrained by hermiticity, parity and time-reversal symmetries~\cite{Meissner:2009ww,Lorce:2013pza,Lorce:2015sqe}
\begin{align}
W_{\lambda\lambda^{\prime}}^{[\G],q}(P,\bk,\Delta,n;\eta)&=
\left[W_{\lambda^{\prime}\lambda}^{[\gamma^0\G^\dagger\gamma^0],q}(P,\bk,-\Delta,n;\eta)\right]^*&&\text{hermiticity}\label{hermiticity}\\
&=
W_{\lambda_{\mathsf P}\lambda^{\prime}_{\mathsf P}}^{[\gamma^0\G\gamma^0],q}(P_{\mathsf P},\bk_{\mathsf P},\Delta_{\mathsf P},n_{\mathsf P};\eta)&&\text{parity}\\
&=
\left[W_{\lambda_{\mathsf T}\lambda^{\prime}_{\mathsf T}}^{[(i\gamma_5C)\G^*(i\gamma_5C)],q}(P_{\mathsf T},\bk_{\mathsf T},\Delta_{\mathsf T},n_{\mathsf T};-\eta)\right]^*&&\text{time-reversal}
\label{timereversal}
\end{align} 
where $\eta=+1$ (-1) in DIS (Drell-Yan) kinematics, and the labels $\mathsf P$ and $\mathsf T$ indicate that the variables are transformed by parity and time-reversal, respectively. 
We stress in particular that the $\xi$-dependence is constrained by the hermiticity property.

The GTMD $W_{\lambda\lambda^{\prime}}^{[\g^+],q}$ can be decomposed as follows~\cite{Meissner:2009ww}
\begin{equation}
\label{eq:ParameterizationGTMDs}
W_{\lambda\lambda^{\prime}}^{[\g^+],q}=\Gamma_{1}\, F^q_{1,1}+\Gamma_{2}\, F^q_{1,2}+\Gamma_{3}\, F^q_{1,3}+\Gamma_{4}\, F^q_{1,4}
\,,
\end{equation}
where the $F^q_{1,i}$ functions are in general complex-valued and the Dirac helicity structures $\G_i$ are~\footnote{Strictly speaking, the variables $p,p',P,\l,\l'$ refer to partonic variables in all the results below. However we use the same symbols as for hadronic variables.}
\begin{align}\label{eq:StructuresGamma14}
\Gamma_{1} & = 
\frac{1}{2M}\bar{u}(p^{\prime},\lambda^{\prime})u(p,\lambda)
\,,\qquad\qquad
&&\hspace{-3cm}
\Gamma_{2} =
\frac{1}{2M}\frac{\vecb\bk_{\perp}^{i}}{P^{+}}\,
\bar{u}(p^{\prime},\lambda^{\prime})i\sigma^{i+}u(p,\lambda)
\,,\nonumber \\
\Gamma_{3} & =
\frac{1}{2M}\frac{\vecb\D_{\perp}^{i}}{P^{+}}\,
\bar{u}(p^{\prime},\lambda^{\prime})i\sigma^{i+}u(p,\lambda)
\,,\qquad\qquad
&&\hspace{-3cm}
\Gamma_{4} = 
\frac{1}{2M}\frac{\vecb \bk_{\perp}^{i}\vecb\D_{\perp}^{j}}{M^{2}}\,
\bar{u}(p^{\prime},
\lambda^{\prime})i\sigma^{ij}u(p,\lambda)
\,.
\end{align}
These structures have been chosen so that the real (imaginary) part of the $F^q_{1,i}$ functions is $\eta$-even (odd).

At tree level the soft function is unity, as well as the Wilson lines in the collinear correlator. 
Thus we simply have
\begin{eqnarray}\label{eq:WLO}
W_{\lambda\lambda^{\prime}}^{[\g^+],q}\Big|_{\mathrm{LO}} & = & 
\frac{1}{2P^{+}}\delta(1-x)\, \delta^{(2)}(\vecb\bk_{\perp})\,
\bar{u}(p^{\prime},\lambda^{\prime})\g^+u(p,\lambda)
+\mathcal{O}(\alpha_{s})
\,.
\end{eqnarray}
Comparing this result with the general parametrization \eqref{eq:ParameterizationGTMDs}, we can identify the various $F_{1,i}$ distributions (using Gordon identities that appear, for instance, in Appendix~A of \cite{Meissner:2009ww}):
\begin{eqnarray}\label{eq:F1iLO}
F_{1,1}^{q} & = & (1-\xi^{2})\delta(1-x)\,\delta^{(2)}(\vecb\bk_{\perp})+\mathcal{O}(\alpha_{s})
\,,\nonumber \\
F_{1,2}^{q} & = & \mathcal{O}(\alpha_{s})
\,,\nonumber \\
F_{1,3}^{q} & = & \frac{1}{2}\delta(1-x)\,\delta^{(2)}(\vecb\bk_{\perp})+\mathcal{O}(\alpha_{s})
\,,\nonumber \\
F_{1,4}^{q} & = & \mathcal{O}(\alpha_{s})
\,.
\end{eqnarray}

Let us now consider the NLO corrections, starting from the virtual ones.
The soft function has already been calculated with the $\d$-regulator in \cite{GarciaEchevarria:2011rb}~\footnote{Please note that the soft function contribution to the quark GTMD is the same as for semi-inclusive DIS (SIDIS). The latter process is our choice for the underlying kinematics. Due to universality of the soft function, this choice is quite general and does not affect the properties of the GTMDs.}.
Using the LO result for the unsubtracted matrix element $\phi$ of \eqref{eq:GTMDPDFDef}, which is basically $\delta(1-x)$, we can express the NLO contribution of the soft function to the GTMD in (\ref{eq:GTMDPDFDef}).
It is
\begin{align}\label{eq:WSFvirt}
&W_{\lambda\lambda^{\prime}}^{[\g^+],q}\Big|_{S\,\mathrm{virtual}} =  
\frac{1}{2}\left \{
\frac{1}{2P^{+}}\bar{u}(p^{\prime},\lambda^{\prime})\g^+u(p,\lambda)\,
\delta(1-x)\,\delta^{(2)}(\vecb\bk_{\perp})\,
\frac{\alpha_{s}C_{F}}{2\pi}
\Bigg[-\frac{2}{\veuv^{2}}
+\frac{2}{\veuv}\ln\frac{\delta^{2}}{\mu^{2}}
-\ln^{2}\frac{\delta^{2}}{\mu^{2}}
+\frac{\pi^2}{2}\Bigg] \right \}
\,,
\end{align}
where the overall factor of $1/2$ comes from the square root of $S$ in \eqref{eq:GTMDPDFDef}.
This factor will be taken into account below. One can already notice the term of mixed UV-RD divergences $\frac{1}{\veuv} \ln \d$. 
This term exemplifies one form through which the problematic feature of RDs is manifested, namely their entanglement with UV divergences. 
Moreover this term does not cancel when real gluon contribution from the soft function is included. 
This is simply because the TMD soft function restricts the transverse momentum of the emitted real gluons to be finite.

The virtual gluon contribution from the collinear matrix element $\phi$ is

\begin{align}\label{eq:PhiVirt}
&W_{\lambda\lambda^{\prime}}^{[\g^+],q}\Big|_{\phi\,\mathrm{virtual}} = 
\frac{1}{2P^{+}}\bar{u}(p^{\prime},\lambda^{\prime})\g^+u(p,\lambda)\,
\delta(1-x)\,\delta^{(2)}(\vecb\bk_{\perp})\,
\frac{\alpha_{s}C_{F}}{2\pi}\,
\Bigg[\frac{2}{\veuv}\ln\frac{\delta}{P^{+}\sqrt{1-\xi^{2}}} 
+ \frac{3}{2\veuv}
\nonumber \\
&\quad
-\frac{3}{2}\ln\frac{\tilde\Delta}{\mu^{2}} 
- 2\ln\frac{\tilde\Delta}{\mu^{2}} \ln\frac{\delta}{P^{+}\sqrt{1-\xi^{2}}}
-\frac{1}{2}\ln^{2}\frac{\delta}{P^{+}(1+\xi)}
-\frac{1}{2}\ln^{2}\frac{\delta}{P^{+}(1-\xi)}
+\frac{7}{4}+\frac{5}{12}\pi^{2}
{+i\pi\ln\frac{1-\xi}{1+\xi}}
\Bigg]
\,,
\end{align}
where the last result \emph{still} includes the soft contamination (and thus it is known in the literature as the \emph{naive} contribution).  
It also clearly shows the presence of mixed divergences $\frac{1}{\veuv}\ln\d$. 
Notice that the result in \eqref{eq:PhiVirt} reduces to the analogous one for the TMDPDF (denoted by ${\hat f}_{n1}^{v,DIS}$ in (6.7) in \cite{GarciaEchevarria:2011rb}), once the forward limit and the averaging over polarizations are performed.
We also draw the attention to the imaginary term, which is  consistent with the hermiticity property~\eqref{hermiticity}. 
This term involves also an implicit $\eta$ dependence (equal to $+1$ in our DIS kinematics) owing to time-reversal symmetry~\eqref{timereversal}. 
We remark that this structure did not appear in~\cite{Lorce:2015sqe} because the considered distributions were defined for $\xi=0$.
In the model calculations performed in \cite{Kanazawa:2014nha,Mukherjee:2014nya}, one can notice divergences for $\xi = 0$.
Those divergences echo the notion of rapidity divergences encountered above.

To calculate all the virtual contributions to the GTMD we need to subtract a whole contribution of the soft function, thus obtaining the pure collinear, and then add half of it as dictated by \eqref{eq:GTMDPDFDef}. 
We thus end up subtracting \eqref{eq:WSFvirt} from \eqref{eq:PhiVirt}. 
The result is

\begin{align}\label{eq:Wvirtual}
&W_{\lambda\lambda^{\prime}}^{[\g^+],q}\Big|_{\mathrm{virtual}} = 
\frac{1}{2P^{+}}\bar{u}(p^{\prime},\lambda^{\prime})\g^+u(p,\lambda)\,
\delta(1-x)\,\delta^{(2)}(\vecb\bk_{\perp})\,
\frac{\alpha_{s}C_{F}}{2\pi}
\Bigg[\frac{1}{\veuv^{2}}
+\frac{1}{\veuv}\left(\frac{3}{2}+\ln\frac{\mu^{2}}{Q^2(1-\xi^{2})}\right)
\nonumber \\
&
-\frac{3}{2}\ln\frac{\tilde\Delta}{\mu^{2}}
-\frac{1}{2}\ln^{2}\frac{\tilde\Delta^{2}}{\mu^{2}Q^2}
+\ln^{2}\frac{\tilde\Delta}{\mu^{2}}
+\ln\frac{\tilde\Delta^2}{\mu^2Q^2}\ln(1-\xi^{2})
-\frac{1}{2}\ln^{2}(1+\xi)-\frac{1}{2}\ln^{2}(1-\xi)
+\frac{7}{4}+\frac{\pi^{2}}{6}
{+i\pi\ln\frac{1-\xi}{1+\xi}}
\Bigg]
\,,
\end{align}
where $Q^2=P^+P^-$, and without loss of generality we have used the relation $\tilde\D\equiv Q\d$ with $Q=P^+=P^-$ to simplify the logarithmic structure.
It is clear that the result \eqref{eq:Wvirtual} is free from mixed UV-RDs, as anticipated.
Moreover the last result is consistent with the one of quark TMDPDF (denoted by $j_{n1}^{v,DIS}$ in (6.7) in \cite{GarciaEchevarria:2011rb}).

Now, regarding the real-gluon emission diagrams, we have calculated their contribution to the GTMD following the same steps as for the virtual part, and full details will be given elsewhere. 
When all contributions are added together, we get the complete result for the GTMD, $W_{\lambda\lambda^{\prime}}^{[\g^+],q}$, at NLO.
In the following we set both $\tilde\D=0$ and $\eps=0$ unless they regulate any divergence.
For the four structures appearing in \eqref{eq:ParameterizationGTMDs} we have, in the DGLAP region $|\x|<x$,

\begin{align}\label{eq:F11}
&F_{1,1}^{q}(|\xi|<x) = 
(1-\xi^{2})\,\delta(1-x)\,\delta^{(2)}(\vecb\bk_{\perp})
\Bigg\{1+\frac{\alpha_{s}C_{F}}{2\pi}
\Bigg[\frac{1}{\veuv^{2}}
+\frac{1}{\veuv}\left(
\frac{3}{2}+\ln\frac{\mu^{2}}{Q^2(1-\xi^{2})}\right)
\nonumber\\
&\quad
-\frac{3}{2}\ln\frac{\tilde\Delta}{\mu^{2}}
-\frac{1}{2}\ln^{2}\frac{\tilde\Delta^2}{\mu^{2}Q^2}
+\ln^{2}\frac{\tilde\Delta}{\mu^{2}}
+\ln\frac{\tilde\Delta^2}{\mu^{2}Q^2}\ln(1-\xi^{2})
-\frac{1}{2}\ln^{2}(1+\xi)
-\frac{1}{2}\ln^{2}(1-\xi)
+\frac{7}{4}+\frac{\pi^{2}}{6}
+i\pi \ln\frac{1-\xi}{1+\xi}
\Bigg]\Bigg\}
\nonumber\\
&\quad
+\frac{\alpha_{s}C_{F}}{2\pi^{2}}\Bigg[
\frac{(1-\xi^{2})N_{1}-\frac{\D_{T}^{2}}{2}N_{2}+\xi\vecb\bk_\perp\cd\vecb\D_\perp N_{3}}
{D_{+}D_{-}} 
- (1-\xi^{2})\,\delta(1-x)
\frac{1}{\bk_{T}^{2}}\ln\frac{\bk_{T}^{2}}{Q^2}\Bigg]
\,,
\end{align}
where the functions $N_i$ and $D_\pm$ are
\begin{eqnarray}
N_{1} & = & 
\frac{1}{(1-\xi^{2})(1-x)_+}\Bigg[
\left((1+x^{2})-2\xi^{2}\right)\bk_{T}^{2}
+2\xi(1-x)\vecb\bk_{\perp}\cd\vecb\D_{\perp}
-(1-x)^{2}\frac{\D_{T}^{2}}{2}
\Bigg]
\,,\nonumber \\
N_{2} & = & -\frac{(1-x)^{2}}{2(1-\xi^{2})}\left(1+x\right)
\,,\quad\quad
N_{3}  =  -\frac{1-x}{1-\xi^{2}}\left(1+x\right)
\,,\quad\quad
D_{\pm}  =  \left(\vecb\bk_\perp\pm\frac{1-x}{1\mp\xi}\frac{\vecb\D_\perp}{2}\right)^{2}
\,.
\end{eqnarray}
We notice that terms proportional to $\eps$ were dropped in the functions $N_i$.
The other distributions are
\begin{eqnarray}\label{eq:F12}
F_{1,2}^{q}(|\xi|<x) & = & 
\frac{\alpha_{s}C_{F}}{2\pi^{2}}
\Bigg[\frac{\frac{\xi\D_{T}^{2}}{2(1-\xi^{2})}N_{3}}{D_{+}D_{-}}\Bigg]
\,,
\end{eqnarray}
\begin{align}\label{eq:F13}
&F_{1,3}^{q}(|\xi|<x) = 
\frac{1}{2}\,\delta(1-x)\,\delta^{(2)}(\vecb\bk_{\perp})
\Bigg\{1+\frac{\alpha_{s}C_{F}}{2\pi}
\Bigg[\frac{1}{\veuv^{2}}
+\frac{1}{\veuv}\left(
\frac{3}{2}+\ln\frac{\mu^{2}}{Q^2(1-\xi^{2})}\right)
\nonumber\\
&\quad
-\frac{3}{2}\ln\frac{\tilde\Delta}{\mu^{2}}
-\frac{1}{2}\ln^{2}\frac{\tilde\Delta^2}{\mu^{2}Q^2}
+\ln^{2}\frac{\tilde\Delta}{\mu^{2}}
+\ln\frac{\tilde\Delta^2}{\mu^{2}Q^2}\ln(1-\xi^{2})
-\frac{1}{2}\ln^{2}(1+\xi)
-\frac{1}{2}\ln^{2}(1-\xi)
+\frac{7}{4}+\frac{\pi^{2}}{6}
+i\pi \ln\frac{1-\xi}{1+\xi}
\Bigg]\Bigg\}
\nonumber\\
&\quad
+\frac{\alpha_{s}C_{F}}{2\pi^{2}}\Bigg[
\frac{\frac{1}{2}N_{1}-\frac{\D_{T}^{2}}{4(1-\x^2)}N_{2}}
{D_{+}D_{-}} 
- \frac{1}{2}\,\delta(1-x)
\frac{1}{\bk_{T}^{2}}\ln\frac{\bk_{T}^{2}}{Q^2}\Bigg]
\,,
\end{align}
and
\begin{eqnarray}\label{eq:F14}
F_{1,4}^{q}(|\xi|<x) & = & 
\frac{\alpha_{s}C_{F}}{2\pi^{2}}\Bigg[\frac{-M^{2}N_{3}}{D_{+}D_{-}}\Bigg]
\,. 
\end{eqnarray}

These results are consistent with the constraints~\eqref{hermiticity}-\eqref{timereversal}.  In particular, real terms involving transverse momenta are necessarily functions of $\bar k_T^2$, $\Delta^2_T$ and $\xi\vecb\bk_\perp\cd\vecb\D_\perp$, since momentum transfer has to appear with an even power.
It should me mentioned that in these results we considered the leading-power contributions in $M^2$.

In the expressions for $F_{1,1}^q$ and $F_{1,3}^q$ we notice the appearance of single and double logarithms of ${\tilde{\Delta}}$. 
When transforming to coordinate space all the double logarithms will cancel. 
For that to happen one has to retain all the $\delta$ dependence in the last line in \eqref{eq:F11}, which was dropped because it is not needed in momentum space. Details of this calculation will be presented elsewhere. 
The origin of the double logarithms in ${\tilde{\Delta}}$ can be traced to rapidity divergences, and they appear in real and virtual contributions for both unsubtracted collinear and soft matrix elements. 
The cancellation of such double logarithms is crucial to obtain the evolution kernel in an exactly similar manner as the case for TMDs. 
The remaining single logarithm in ${\tilde{\Delta}}$ in $F_{1,1}^q$ signals a genuine long-distance effect. 
For the simpler case of unpolarized TMDPDF, this single logarithm is exactly the collinear divergence in the integrated PDF (when the $\delta$ regulator is implemented)~\cite{GarciaEchevarria:2011rb}.

To conclude this Section, we observe that the inclusion of the soft function in the definition of GTMDs completely cancels spurious rapidity divergences and makes them well-defined hadronic quantities, with a proper evolution (discussed below) and all the properties that one would expect to have in such objects.

\section{Evolution of GTMDs}
\label{sec:evolution}

The evolution of GTMDs is identical to the one of TMDs, since both quantities are defined through the same bi-local operator (and the same soft function).
Thus GTMDs also depend on two scales, the renomalization scale $\m$ and rapidity scale $Q^2$.
Since both quark and gluon GTMDs have an analogous operator structure, in this Section we discuss the evolution of both types of distributions at once.

The evolution in $\m$ is governed by the anomalous dimension $\g_W^{j}$ (for $j=q,g$),
\begin{align}
\frac{d}{d\ln\m} \ln\tilde W^{j}(b_T;\m,Q^2) &=
\g_W^{j}\(\as(\m),\ln\frac{Q^2(1-\x^2)}{\m^2}\)
\,,
\end{align}
where $\tilde W^{q}$ represents, at leading twist, the transform of any of the 16 quark GTMDs that parametrize $W_{\lambda\lambda^{\prime}}^{[\G],q}$ \cite{Meissner:2009ww} to coordinate space, with $b_T$ being the conjugate variable of $\bk_T$ (a similar discussion applies to $\tilde W^{g}$ \cite{Lorce:2013pza}).
$\g_W^j$ has the same functional form as the anomalous dimension of the TMDs (see e.g. \cite{Echevarria:2012pw} for the quark case and \cite{Echevarria:2015uaa} for the gluon case),
\begin{align}
\g_W^j\(\as(\m),\ln\frac{Q^2(1-\x^2)}{\m^2}\) &=
-\G_{\rm cusp}^{j}\big(\as(\m)\big)\ln\frac{Q^2(1-\x^2)}{\m^2} - \g^j\big(\as(\m)\big)
\,.
\end{align}
The cusp anomalous dimension $\G_{\rm cusp}^j$ is taken either in the fundamental or adjoint representation, depending on whether we are considering  quark or gluon GTMDs, respectively.
The non-cusp piece $\g^j$ also depends on whether we are considering quark or gluon GTMDs (it corresponds to $\g^V$ in \cite{Echevarria:2012pw} for quarks and $\g^{nc}$ in \cite{Echevarria:2015uaa} for gluons).
 All these anomalous dimensions are currently known up to third order in $\alpha_s$.

The evolution in $Q^2$ is given by
\begin{align}
\frac{d}{d\ln Q^2} \ln \tilde W^{j}(b_T;\m,Q^2) &=
-D^j(b_T;\m)
\,,
\end{align}
where the $D^j$ function is the same as for the TMDs, since, for both classes of hadronic quantities, it is obtained from the contribution of the soft function, which is the same in both. 
See \cite{Echevarria:2012pw,Echevarria:2015byo,Echevarria:2015uaa} for the role of the $D^j$ term in the evolution of TMDs.

Combining the dependence of GTMDs on both scales, their complete evolution from a given initial scales $(\m_0,Q_0^2)$ to some final scales $(\m,Q^2)$ is given by
\begin{align}
\tilde W^{j}(b_T;\m,Q^2) &=
R^j(\x,b_T;\m,Q^2,\m_0,Q_0^2)\,
\tilde W^{j}(b_T;\m_0,Q_0^2)
\,,
\end{align}
where the evolution kernel $R^j$ is
\begin{align}
R^j(\x,b_T;\m,Q^2,\m_0,Q_0^2) &=
\(\frac{Q^2}{Q_0^2}\)^{-D^j(b_T;\m_0)}\,
\exp\[\int_{\m_0}^{\m} \frac{d\hat\m}{\hat\m}\,
\g_W^j\(\as(\hat\m),\ln\frac{Q^2(1-\x^2)}{{\hat\m}^2}\)\]
\,.
\end{align}
The $D^j$ term can be calculated perturbatively only in the small $b_T$ region, and thus one needs to parametrize the large $b_T$ tail with some non-perturbative model.
We emphasize the fact that this model has to be the same as the one for TMDs, since the soft function that enters the definitions of both TMDs and GTMDs is the same. Below we exploit this universality to illustrate the impact of evolution on $F_{1,1}^q$.

To simplify our presentation we consider the quark GTMD $F^q_{1,1}$ at an input scale $Q_0=\m_0$ and zero skewness ($\x=0$), and model it with a simple and factorized gaussian dependence on the transverse momentum:
\begin{align}\label{eq:model1}
F^q_{1,1}(x,\x=0,\bk_T^2,\D_T^2,\vecb\bk_\perp\cd\vecb\D_\perp;Q_0) = 
H^q(x,\x=0,t=-\D_T^2;Q_0)\,
\frac{e^{-\bk_T^2/\mean{\bk_T^2}}}{\pi\mean{\bk_T^2}}
\,,
\end{align}
where $\mean{\bk_T^2}$ stands for the width of the GTMD and $H^q$ is the unpolarized quark GPD.
For the latter, and as an  ansatz, we parametrize it as a product of its corresponding collinear distribution in the forward limit and a $t$-dependent function:
\begin{align}\label{eq:model2}
H^q(x,\x=0,t;Q_0) = f_1^q(x;Q_0)\, \exp\big[\l\,t\big]
\,,
\end{align}
with $\l=0.5\,{\rm GeV}^{-2}$.
For recent phenomenological works on GPDs see e.g. \cite{Favart:2015umi,Guidal:2013rya}.

Finally, for the $D^q$ term we choose the following implementation \cite{Echevarria:2012pw}
\begin{align}
D^q(b_T;Q_0) &=
D^q(b_T^*;\m_b) + \int_{\m_b}^{Q_0}\frac{d\bar\m}{\bar\m} \G_{\rm cusp}^q
+ \frac{1}{4} g_2 b_T^2
\,,\qquad
b_T^* = b_T\[1+\(\frac{b_T}{b_{\rm max}}\)^2\]^{-\frac{1}{2}}
\,,\qquad
\m_b = 2e^{-\g_E}/b_T^*
\,.
\end{align}

\begin{figure}[t]
\begin{center}
\includegraphics[width=0.4\textwidth]{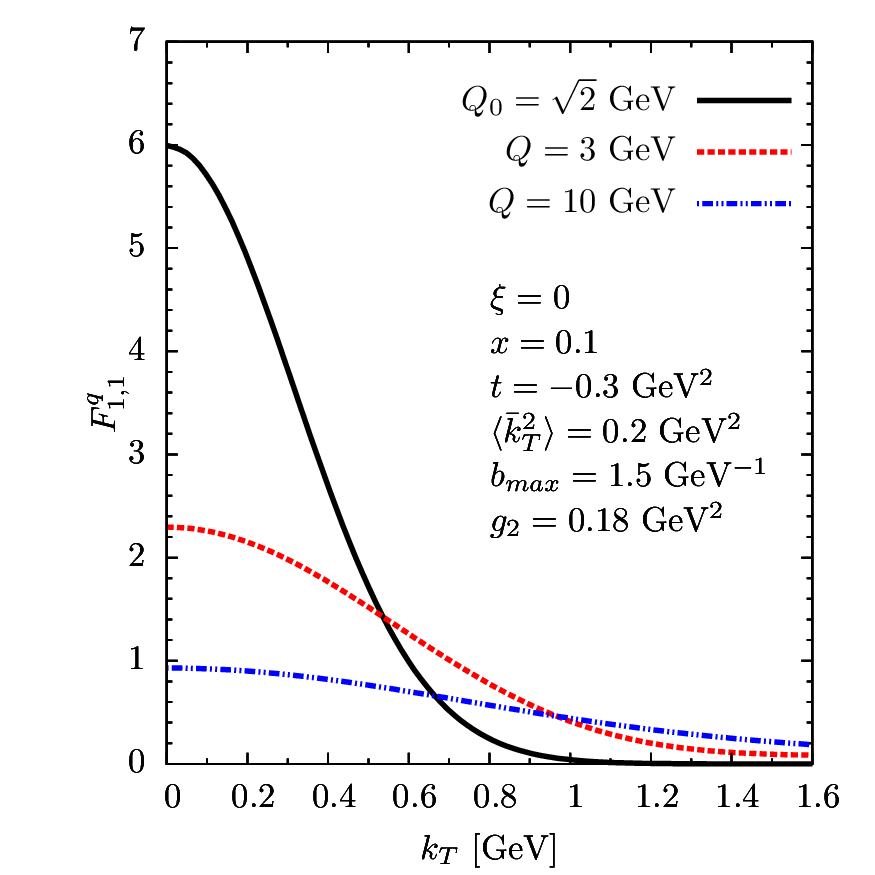}
\end{center}
\caption{\it
Evolved GTMD $F_{1,1}^q$ from initial scale $Q_0=\sqrt{2}~{\rm GeV}$.
}
\label{fig:plot_gtmd}
\end{figure}

In Fig.~\ref{fig:plot_gtmd} we show the GTMD at the input scale for a particular choice of the variables and parameters, and the effect of evolution when it is evolved up to another two scales (with $Q=\m$).
Evolution is implemented at next-to-next-to-leading-logarithmic accuracy, which needs $\G_{\rm cusp}^q$ at three loops, $\g^q$ at two loops and $D^q$ at two loops; they can be found in \cite{Echevarria:2012pw}.
We also used the MSTW08nnlo set for the unpolarized collinear PDF $f_1^q$ \cite{Martin:2009iq}.
As already mentioned, the non-perturbative input needed for the $D^q$ term in the evolution kernel is the same as the one needed for TMDs.
In this case we use the model found in \cite{Konychev:2005iy}.
As can be clearly noticed, the evolution flattens and widens the distribution. 
The same effect was observed for TMDs as well (see e.g. \cite{Echevarria:2012pw,Aybat:2011zv}), which is of course not surprising in view of the simple model for the GTMD $F_{1,1}^q$ in \eqref{eq:model1}-\eqref{eq:model2}.

\section{Conclusions}
\label{sec:conclusions}

In this work we have considered the current formulation of generalized transverse momentum dependent distributions (GTMDs). 
We argued that this formulation leads to ill-defined quantities. 
This observation relied mainly on our understanding of how a three-dimensional ($1+2$) formulation of hadronic matrix elements ought to be constructed. 
This observation was supported by a first-order calculation in perturbative QCD.
The latter demonstrated the appearance of the anticipated spurious rapidity divergences and their cancellation once the proper definition is considered, by the inclusion of a non-trivial soft gluon radiation factor. 
The formulation we presented shows explicitly the relation of the newly defined quantities with the current formulation of transverse momentum dependent parton distributions. 
Moreover, it does not alter the relation between GTMDs and generalized parton distributions -- a notion that will be discussed elsewhere.
With the proper definition of the GTMDs at hand, we managed to obtain their evolution kernel. 
As for the case of TMDs, the evolution kernel is spin independent. 
We used the currently known perturbative ingredients to resum its large logarithms up to next-to-next-to-leading-logarithmic accuracy, and illustrated the impact of evolution on the physically interesting function $F^q_{1,1}$.
To do so we have also exploited the fact that the non-perturbative contribution to the evolution kernel of the GTMDs is the same as the one that drives the evolution of TMDs, since in both cases we have the same soft function contribution.
Our findings, demonstrated for quark GTMDs, apply as well to the gluon GTMDs.

\section*{Acknowledgements}

M.G.E. thanks Nikhef, where part of this work was done. 
A.I. thanks John C. Collins for useful discussions.
M.G.E. is supported by the Spanish Ministry of Economy and Competitiveness under the \emph{Juan de la Cierva} program and grant FPA2013-46570-C2-1-P.
This work has been supported by the National Science Foundation under Contract No. PHY-1516088 (K.K. and A.M.).


\end{document}